# Viscoelasticity of model surfactant solutions determined by rotational magnetic spectroscopy

F. Loosli, M. Najm and J.-F. Berret*

*Matière et Systèmes Complexes, UMR 7057 CNRS Université Denis Diderot Paris-VII, Bâtiment Condorcet, 10 rue Alice Domon et Léonie Duquet, 75205 Paris, France.*

**Abstract:** Being able to reduce the size of a rheometer down to the micron scale is a unique opportunity to explore the mechanical response of expensive and/or confined liquids and gels. To this aim, we synthesize micron size wires with magnetic properties and examine the possibility of using them as microrheology probes. In this work, we exploit the technique of rotational magnetic spectroscopy by placing a wire in a rotating magnetic field and monitor its temporal evolution by time-lapse microscopy. The wire-based microrheology technique is tested on wormlike micellar surfactant solutions showing very different relaxation dynamics and viscosities. A model for the wire rotation is also developed and used to predict the wire behavior. It is shown that the rheological parameters of the surfactant solutions including the static shear viscosity, the entangled micellar network relaxation time and the elastic modulus are in good agreement with those of conventional rheometry.

**Keywords**: viscoelastic solids – microrheology – magnetic wires – Maxwell model

## I - Introduction

Rheology is the study of how complex fluids flow and deform under stress [1, 2]. Traditional rheometers measure the frequency-dependent linear viscoelastic relationship between strain and stress on milliliter scale samples. Microrheology in contrast measures these quantities using colloidal probes directly embedded in the fluid [3, 4]. Fluids produced in tiny amounts or confined in small volumes, down to 1 picoliter can be examined by this technique. The past 20 years have seen significant advances in the field, both theoretically and experimentally [5].

In microrheology, the objective is to translate the motion of a probe particle into the relevant rheological quantities such as the elastic complex modulus or the creep response function [6, 7]. Following the work by Crick and Hughes [8], recent studies have shown that microrheology based on the use of anisotropic probes, such as wires, rods or needles could bring notable contributions to the field [5, 9-20]. Anisotropic probes present many advantages compared to spherical ones, in particular these probes behave like conventional stress rheometer when subjected to an external field. It has been found for instance that the static viscosity of a Newtonian fluid can be determined from the motion of micro-actuators submitted to rotating electric or magnetic fields [21-23]. In these studies, the response of the probe over a broad spectral range appears as a resonance peak similar to that found in mechanical systems [24]. This technique was thereafter described as a spectroscopic method for probing fluid viscosity, and dubbed electric or magnetic rotation spectroscopy [25, 26]. Here, we combine this steady rotation technique along with micromechanical response modeling to examine complex fluids displaying both viscosity and elasticity effects. The method is based on the tracking of magnetic wires submitted to a rotational magnetic field at increasing frequencies, and on the analysis of their temporal trajectories [10, 11, 14, 17, 19, 21, 26-29].

In this work, we first introduce a mechanical model for the magnetic wire rotation in Maxwell fluids. Magnetic rotational spectroscopy is thereafter implemented on surfactant wormlike micellar solutions. The solutions investigated are a mixture of cetylpyridinium chloride (CPCl) and sodium salicylate (NaSal) and a mixture of cetyltrimethylammonium bromide (CTAB) and NaSal. Following the work by Rehage and Hoffman [30, 31] CPCl/NaSal and CTAB/NaSal are known to self-assemble spontaneously into micrometer long wormlike micelles and to build semi-dilute entangled networks [32-34]. This network confers to the solution a Maxwell-type viscoelastic behavior [1]. In this study, we compare surfactant solutions that are characterized by different relaxation dynamics and viscosities. It is shown that in such conditions the wire motion exhibit a wide variety of behaviors as a function of the time, including steady rotation, oscillations, continuous and discontinuous back motions. These behaviors depend on a newly defined parameter expressed as the product of the Maxwell relaxation time and critical frequency. The rheological parameters of the surfactant solutions studied are in good agreement with those of conventional rheometry.

## II – Materials and Methods

### II.1 - Iron oxide nanowires

Iron oxide nanoparticles were synthesized by co-precipitation of iron(II) and iron(III) salts in aqueous media and by further oxidation of the magnetite ($Fe_3O_4$) into maghemite ($\gamma$-$Fe_2O_3$) [35, 36]. The particle size and dispersity were determined from transmission electron microscopy (Jeol-100 CX) at $D_{TEM}$ = 13.2 nm and $s_{TEM}$ = 0.23, whereas the maghemite structure was assessed by electron beam diffraction [37]. Light scattering (NanoZS, Malvern) was used to measure the weight-average molecular weight ($M_w$ = 12×10$^3$ kDa) and the hydrodynamic diameter ($D_H$ = 27 nm) of the uncoated particles [35, 36]. The wires were made according to a bottom-up co-assembly process using the $\gamma$-$Fe_2O_3$ particles coated with poly(acrylic) acid together with cationic polymers [37, 38]. The polymer used was poly(diallyldimethylammonium chloride) (PDADMAC, Sigma Aldrich) of molecular weight $M_w$ = 26.8 kDa. Determined by size exclusion chromatography, the degree of polymerization and the dispersity Đ were found equal to 50 and to 3.5 respectively. The wires dispersion was autoclaved at 120 °C and atmospheric pressure during 20 minutes to prevent bacterial contamination and stored at 4 °C.

### II.2 – Maxwell viscoelastic solutions

The surfactant solutions investigated were a mixture of cetylpyridinium chloride and sodium salicylate at a concentration of 2 wt. % in 0.5M NaCl [30, 33, 34] and a mixture of cetyltrimethylammonium bromide and sodium salicylate at c = 1 wt. % [31]. At the concentration of 1 - 2 wt. %, the network mesh size is of the order of 30 nm, *i.e.* much smaller than the wire diameter [33]. A cone-and-plate and controlled shear rate rheometer (diameter 50 mm, MCR 500 Physica) was used to determine the frequency dependence of the elastic complex modulus $G^*(\omega) = G'(\omega) + iG''(\omega)$. At the temperature T = 27 °C, the two wormlike micellar fluids studied are almost perfect Maxwell fluids, *i.e.* characterized by a unique relaxation time. CPCl/NaSal at 2 wt. % is associated with a static viscosity $\eta_0$ = 1.0 ± 0.1 Pa s, a relaxation time $\tau$ = 0.14 ± 0.01 s and an elastic modulus $G$ = 7.1 ± 0.1 Pa, whereas CTAB/NaSal 1 wt. % is given with $\eta_0$ = 40 ± 4 Pa s, $\tau$ = 23 ± 3 s and an elastic modulus $G$ = 1.7 ± 0.2 Pa. The two fluids have hence a similar entangled network structure, but differ in their micellar network dynamics. The breaking and recombination time $\tau$ for CTAB/NaSal is around 160 times larger than that of CPCl/NaSal.

### II.3 - Microrheology and electromagnetic coils device

Bright field microscopy was used to monitor the wire actuation as a function of time. Stacks of images were acquired on an IX73 inverted microscope (Olympus) equipped with a 100× objective. For magnetic rotation spectroscopy experiments, 65 µl of surfactant solution were deposited on a glass plate and sealed into to a Gene Frame® (Abgene/Advanced Biotech) dual adhesive system. The glass plate was introduced into a homemade device generating a rotational magnetic field, thanks to two pairs of coils (23 Ω) working with a 90°-phase shift (Fig. 1a). An electronic set-up allowed measurements in the frequency range 10$^{-3}$ - 100 rad s$^{-1}$ and at magnetic fields B = 0 – 20 mTesla. Fig. 1b displays the magnetic field distributions between the poles of the electromagnetic coils in the X- and Y-directions. The image acquisition system consisted of an EXi Blue CCD camera (QImaging) working with Metaview (Universal Imaging). Images were digitized and treated by the ImageJ software and plugins. Fig. 1c shows snapshots of a rotating 10 μm wire in a 85 wt. % water-glycerol (Aldrich) mixture at fixed time interval during a 180° rotation. For the wire magnetic property calibration, experiments were performed on a 85 wt. % water-glycerol mixture of static viscosity $\eta_0$ = 0.062 Pa s$^{-1}$ (T = 32 °C).

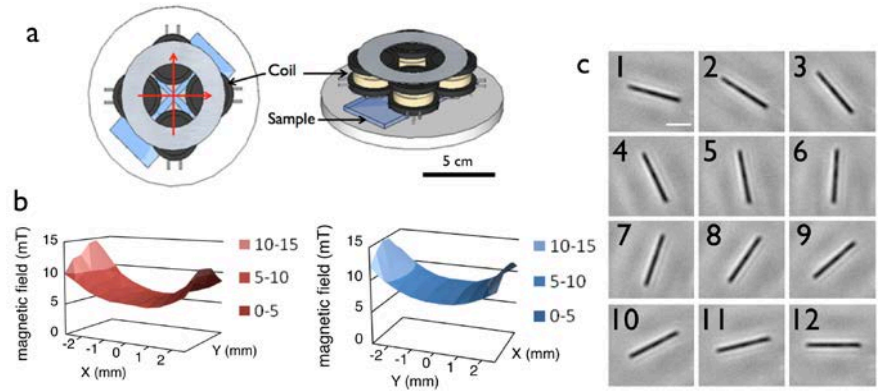

*Figure 1:* a) Top and side views of the rotating field device used in this study. b) Magnetic field distributions are shown along the X and Y-direction of a four-coil device of resistance 23 Ω. In the center, the magnetic field is constant over a surface of 1×1 mm². c) Snapshots of a 10 μm rotating wire. The time scale between two successive images is 46 ms.

## III – Results and Discussion

### III.1 – Modeling wire rotation in Newtonian and Maxwell fluids

*Newtonian liquid*

In continuum mechanics, a Newtonian fluid of viscosity $\eta_0$ is described by a dashpot [2]. In such a fluid, a wire submitted to a rotating field experiences a restoring torque that slows down its motion. With increasing frequency, the wire undergoes a transition between a synchronous and an asynchronous regime. The critical frequency $\omega_C$ between these two regimes reads [21-23]:

$$\omega_C = \frac{3}{8}\frac{\mu_0 \Delta\chi}{\eta_0} g\left(\frac{L}{D}\right)\frac{D^2}{L^2}H^2 \qquad (1)$$

where $\mu_0$ is the permeability in vacuum, $L$ and $D$ the wire length and diameter, $H$ the magnetic excitation amplitude and $g(L/D) = ln(L/D) - 0.662 + 0.917 D/L - 0.050(D/L)^2$ is a dimensionless function of the anisotropy ratio [39]. In Eq. 1, $\Delta\chi = \chi^2/(2+\chi)$ where $\chi$ denotes the magnetic susceptibility. For the analysis, the rotation angle $\theta(t)$ describing the wire in-plane motion is derived and are translated into a set of two parameters: the average rotation velocity $\Omega(\omega)$ for both regimes and the oscillation amplitude $\theta_B(\omega)$ in the asynchronous regime [40]. The average angular velocity $\Omega(\omega)$ expresses as:

$$\begin{aligned}\omega \leq \omega_C &\qquad \Omega(\omega) = \omega \\ \omega \geq \omega_C &\qquad \Omega(\omega) = \omega - \sqrt{\omega^2 - \omega_C^2}\end{aligned} \qquad (2)$$

In Eq. 2, $\Omega(\omega)$ increases linearly with the frequency, passes through a peaked maximum at $\omega_C$ before decreasing. The transition between the synchronous and asynchronous regimes is used for calibration to determine the susceptibility parameter $\Delta\chi$ in Eq. 1. For wires made from 13.2 nm particles and from PADADMAC polymers, we found $\Delta\chi$ = 2.1 ± 0.4, and a magnetic susceptibility $\chi$ = 3.4 ± 0.4 [40].

*Maxwell fluid*

A Maxwell fluid is described by a spring and dashpot in series [2]. An actuated wire immersed in such a medium experiences a viscous and an elastic torque that both oppose the magnetic torque. The main features of the wire rotation are illustrated in Figs. 2 and 3. Fig. 2a shows the time evolution of the rotation angle $\theta(t)$ at frequencies below $\omega_C$, whereas Fig. 2b and 2c display the same quantity above $\omega_C$. The straight line in red indicates the average angular velocity $\Omega$. The two lower figures differ also in the parameter $\theta_0$ defined as:

$$\theta_0 = \lim_{\omega \to \infty} \theta_B(\omega) = \frac{3}{4}\frac{\mu_0 \Delta \chi}{G} g\left(\frac{L}{D}\right)\frac{D^2}{L^2} H^2 \quad (3)$$

The data in Fig. 2b are calculated using $\theta_0$ = 0.2 and those of Fig. 2c for $\theta_0$ = 2. In the regime $\theta_0 < 1$, the decrease during the back motion is a continuous function of the time. In contrast, for $\theta_0 > 1$ the wire rotation displays a mechanical instability which manifests itself by an abrupt jump of the angle. Combining Eqs. 1 and 3, one gets:

$$\theta_0 = 2\omega_C \tau \quad (4)$$

Viscoelastic fluids for which $\theta_0 > 1$ are thus characterized by relaxation times that are larger that half of the inverse critical frequency. Fig. 2d displays the schematic $(\tau,\omega)$-rotation diagram together with the aforementioned regimes. To our knowledge, the regime of long relaxation times ($\tau \gg 1\,s$) has not been addressed yet using magnetic rotation spectroscopy.

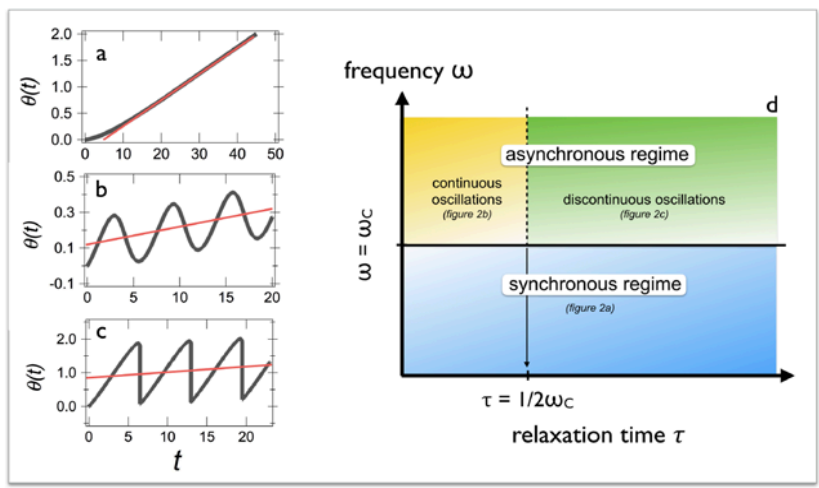

*Figure 2: a-c)* Wire orientation angle $\theta(t)$ in the different rotation regimes below and above the critical frequency $\omega_C$. The red straight lines represent the average angular velocity $\Omega = \overline{d\theta/dt}$. The $\theta(t)$-traces in **b)** and **c)** were obtained for two different values of the elasticity parameter (Eq. 4), $\theta_0$ = 0.2 and 2 respectively. **d)** Schematic $(\tau,\omega)$ diagram of a wire rotating in a Maxwell fluid.

Fig. 3a displays the average rotation frequency evolution $\Omega(\omega)$. Remarkably, the frequency dependence is identical to that of a Newtonian fluid (Eq. 2). The static viscosity $\eta_0$ entering in Eq. 1 is replaced by the product $G\tau$, where $G$ and $\tau$ denote the elastic modulus and the relaxation time. Fig. 3b illustrates the oscillation amplitude $\theta_B(\omega)$ as a function of the reduced frequency. It is calculated for a Newtonian liquid, and for Maxwell fluids at different $\theta_0$-values (0.05, 0.3, 0.8, 1, 2 and 5). Slightly above $\omega_C$, the angle decreases with increasing frequency and then flattens into a frequency independent plateau at $\theta_0$. The plateau in the angle $\theta_B(\omega)$ observed at high frequency is a signature of the medium elasticity. In conclusion, models show that the oscillation amplitude and shape in the asynchronous regime can have a wide variety of behaviors as a function of the time, and that these behaviors depend on the elasticity and on the relaxation time dynamics of the surrounding fluid. In the next sections, it is demonstrated that a quantitative analysis is achievable and that it imparts the correct rheological parameters of the surfactant solutions.

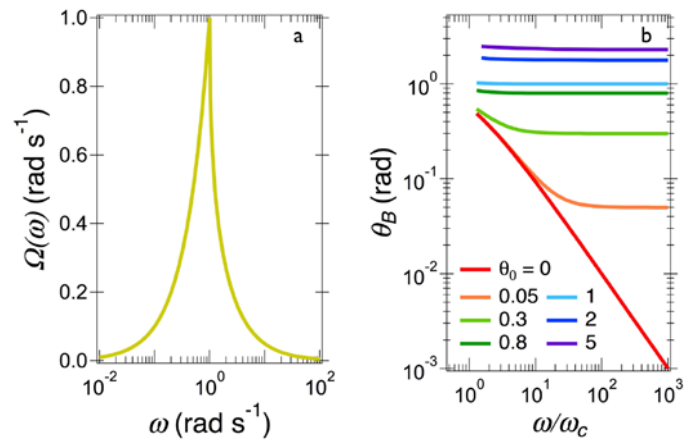

**Figure 3: a)** Average angular velocity $\Omega(\omega)$ as a function of the actuating frequency. The critical frequency is set here at 1 rad s$^{-1}$. **b)** Variation of angle $\theta_B(\omega)$ for Newton and Maxwell models at different values of $\theta_0$ (0, 0.05, 0.3, 0.8, 2 and 5).

### III.2 – Wire rotation transient behavior

In Fig. 4 (upper panels a, b and c), a rotating magnetic field of 15.3 mT was applied to a 8.1µm nanowire immersed in the CPCl/NaSal solution at increasing frequencies, $\omega$ = 0.80, 1.2 and 7.0 rad s$^{-1}$, respectively. The wire motion was monitored by optical microscopy, and their orientation time dependence was derived. At low frequency, the wire rotates with the field and $\theta(t) = \omega t$. Above a critical value, here $\omega_C$ = 0.88 rad s$^{-1}$, the wire is animated by back-and-forth motion characteristic of the asynchronous regime. In this range, $\theta(t)$ displays continuous oscillations. For this solution, the elasticity related parameter is estimated at $\theta_0$ = 0.25 (Eq. 4), which is below the limit of 1 discussed previously. The three lower panels in Fig. 4 (d, e and f) provide the $\theta(t)$-traces for a 15.7 µm wire dispersed in the CTAB/NaSal surfactant solution at frequencies $\omega$ = 0.02, 0.45 and 4.0 rad s$^{-1}$, respectively. Above the critical frequency found at $\omega_C$ = 0.07 rad s$^{-1}$, the wire exhibits sawtooth oscillations, where the back motions are rapid and discontinuous on the time scale considered.

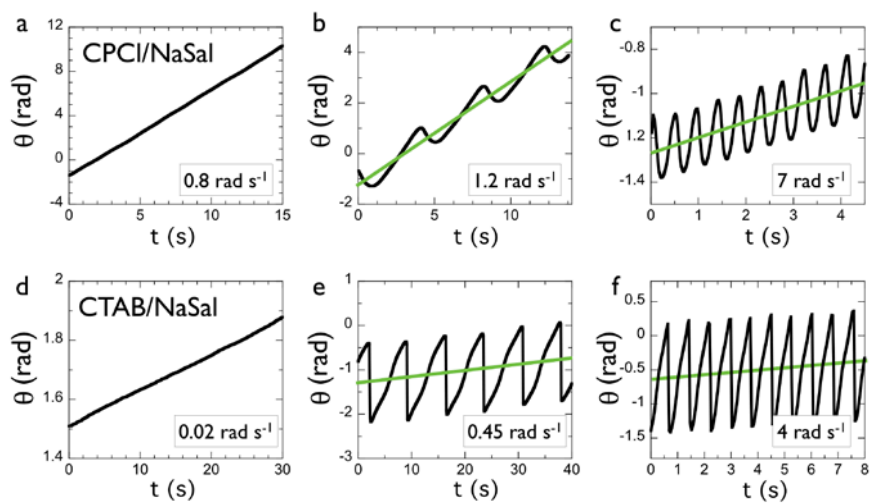

**Figure 4:** Rotation angle $\theta(t)$ as a function of the time for magnetic wires dispersed in CPCl/NaSal (a,b,c) and CTAB/NaSal (d,e,f) surfactant solutions at increasing frequencies. For frequency higher than the critical frequency $\omega_c$ with is equal to 0.88 rad.s$^{-1}$ and 0.07 rad.s$^{-1}$ respectively, the average angular velocity $\Omega = \overline{d\theta/dt}$ is represented by a straight green line.

For the CTAB/NaSal solution, the parameter $\theta_0$ is estimated at 3.2, that is above the limit of 1 discussed in the previous section. These results are in agreement with the predictions found for surfactant solutions with slow and fast dynamics. In particular, it is shown here that the value of the elasticity related parameter $\theta_0$ plays an important role in the oscillation behavior.

### III.3 – Comparing Macro- and microrheology
*CPCl/NaSal surfactant solution*

Fig. 5a shows the viscosity frequency dependence for CPCl/NaSal at T = 27 °C. At low frequency, $\eta(\omega)$ exhibits first a plateau associated with a static viscosity $\eta_0$ = 1.0 ± 0.1 Pa s. As the sweeping frequency is increased, a decrease of the dynamical viscosity is observed and is well accounted for by the Maxwell model (continuous line). Fig. 5b shows the average angular velocity $\Omega(\omega)$ in semi-logarithmic scale obtained from the transient $\theta(t)$-traces of Fig. 4 ($L$ = 8.1 µm, B = 15.3 mT). With increasing frequency, $\Omega(\omega)$ passes through a maximum at the critical value $\omega_C$ = 0.88 rad s$^{-1}$ before decreasing. Least-square calculations using Eq. 2 provide an excellent fit to the data, and a static viscosity $\eta_0 = 1.3 \pm 0.3$ Pa s. Fig. 5c displays the storage and loss moduli $G'(\omega)$ and $G''(\omega)$ frequency dependence together with the Maxwell model calculations (continuous lines). As expected from the model [2], $G'(\omega)$ and $G''(\omega)$ are found to cross at a modulus $G/2$ and at the angular frequency $\omega = 1/\tau$. The $G'(\omega)$-data are compared to those of microrheology determined from oscillation amplitudes $\theta_B(\omega)$ obtained on two different wires of length 8.1 and 8.2 µm (see columns #6 and #7 in Table I for experimental conditions). The expression used to calculate $G'(\omega)$ is an extension of Eq. 3, namely:

$$G'(\omega) = \frac{3}{4}\frac{\mu_0 \Delta\chi}{\theta_B(\omega)} g\left(\frac{L}{D}\right)\frac{D^2}{L^2} H^2 \quad (5)$$

The agreement between macro and microrheology is poor in the viscous regime ($\omega\tau \ll 1$), as $G'(\omega)$ microrheology data always exceed those of conventional rheology. In the elastic regime ($\omega\tau \gg 1$), the agreement is fair, providing elastic moduli of 6.8 ± 2.5 Pa and 9.5 ± 3.2. These results suggest that in the asynchronous range, the back-and-forth motion is specific of the regime at which the fluid is tested, and that the model predictions are well described. Finally, to test the robustness of the above approach, wire spectroscopy experiments were performed on the same CPCl/NaSal surfactant solution using a total of 9 wires of length between 3 and 15 µm, including the 8.1 and 8.2 µm wires discussed previously.

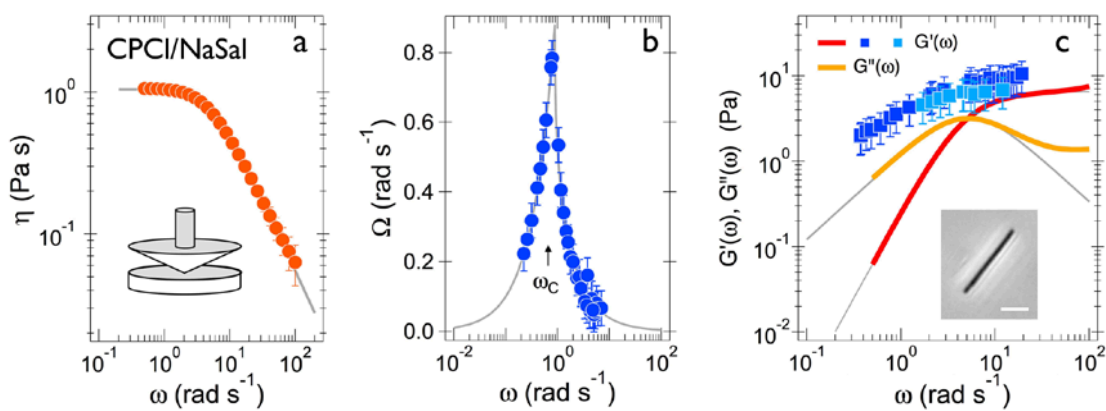

***Figure 5:*** *a) CPCl/NaSal surfactant solution dynamical viscosity at c = 2 wt. % and a temperature of 27 °C. The continuous line is calculated according to the expression $\eta(\omega) = \eta_0/\sqrt{1 + \omega^2\tau^2}$. Inset: schematic representation of a cone-and-plate tool used in conventional rheometry. b) Average angular velocity $\Omega(\omega)$ in semi-logarithmic scale obtained with a 8.1 µm wire actuated under a magnetic field B = 15.3 mT. The continuous line through the data point is obtained using Eq. 2 and $\eta_0 = 1.3 \pm 0.3$ Pa s. c) Elastic and loss moduli $G'(\omega)$ and $G''(\omega)$ measured from cone-and-plate rheometry. The continuous lines in grey are calculated according $G'(\omega) = G\omega^2\tau^2/(1 + \omega^2\tau^2)$ and $G''(\omega) = G\omega\tau/(1 + \omega^2\tau^2)$ with G = 7.1 Pa and $\tau$ =0.14 s. The squares are from microrheology experiments based on the back-and-forth oscillation amplitudes (Eq. 5) for two different wires on length 8.1 and 8.2 µm. Inset: microscopy image of the 8.1 µm magnetic wire.*

For each wire, the critical frequency $\omega_C$ and the oscillation amplitude high frequency limit $\theta_0$ were measured, from which the static viscosity, elastic modulus and relaxation time were derived. Table I displays the experimental conditions used and imparts values for $\eta_0$, $G$ and $\tau$ in accordance with those of cone-and-plate rheometry. Results in Table I indicate that for a better accuracy, it is desirable to repeat the measurement under various experimental conditions.

|  | microrheology | | | | | | | | | rheology |
|---|---|---|---|---|---|---|---|---|---|---|
| wires | #1 | #2 | #3 | #4 | #5 | #6* | #7* | #8 | #9 | cone-and-plate |
| $L$ (μm) | 3.0 | 3.3 | 5.0 | 6.3 | 8.1 | 8.1 | 8.2 | 12.5 | 14.8 | n.d. |
| $B$ (mT) | 14.8 | 14.8 | 14.8 | 10.4 | 10.4 | 15.3 | 10.4 | 7.3 | 10.4 | n.d. |
| $\omega_C$ (rad s$^{-1}$) | 2.1 | 4.0 | 1.6 | 0.80 | 0.38 | 0.88 | 1.44 | 0.41 | 0.55 | n.d. |
| $\eta_0$ (Pa s) | 1.72 | 1.08 | 1.49 | 1.37 | 1.45 | 1.32 | 1.02 | 0.97 | 1.21 | 1.0 |
| $\tau$ (s) | 0.15 | 0.12 | 0.14 | 0.17 | 0.16 | 0.14 | 0.15 | 0.12 | 0.13 | 0.14 |
| $G$ (Pa) | 11.2 | 9.1 | 10.3 | 7.9 | 9.2 | 9.5 | 6.8 | 7.9 | 9.5 | 7.1 |

*Table I:* Viscoelastic parameters of a CPCl/NaSal wormlike micellar solution at 2 wt. % determined from Magnetic Rotational Spectroscopy. The last column shows the values for the static shear viscosity, relaxation time and elastic modulus obtained from cone-and-plate rotational rheology. Wires indicated with a star are related to the data shown in Figs. 5 and 6.

*CTAB/NaSal surfactant solution*
The dynamical viscosity $\eta(\omega)$ obtained for CTAB/NaSal shows a strong decrease on the frequency range tested (Fig. 6a). For slowly relaxing micellar networks such as CTAB/NaSal at 1 wt. %, conventional rheometry provides only a partial view of the rheological response. In particular, the low frequency plateau regime is not accessible and as a result an estimate of $\eta_0$ cannot be inferred from the data. The rheological parameters are actually estimated from the storage and loss modulus asymptotic behaviors instead. Note that for $G''(\omega)$, the fitting using the Maxwell model can only be achieved on a restricted frequency range, reducing the measurement accuracy. Fig. 6b displays the average angular velocity $\Omega(\omega)$ of a 15.7 μm long wire dispersed in CTAB/NaSal ($B$ = 14.8 mT).

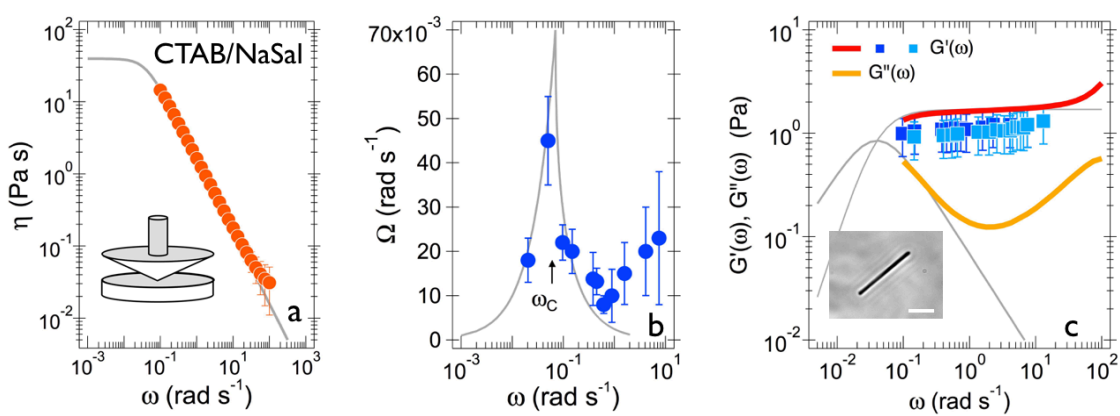

*Figure 6*: Same as in Fig. 5 for a CTAB/NaSal surfactant solution at c = 1 wt. % and a temperature of 27 °C. The continuous lines in a) and c) are obtained using $G$ = 1.7 Pa and $\tau$ = 23 s. The squares in c) are estimated from the oscillation amplitudes in the asynchronous regime for two different wires on length 15.5 and 15.7 μm

The data points exhibit characteristic features similar those discussed previously, and the application of Eq. 2 accounts well for the data. From the resonance peak position, here at $\omega_C$ = 0.07 rad s$^{-1}$, a static viscosity value of $\eta_0$ = 30 ± 8 Pa s is derived. The high frequency $\Omega(\omega)$-data exhibit larger error bars and an increasing tendency, a result that could be related to the existence of the tipping instability similar to that reported by Cappallo and coworkers [27]. As for the storage and loss moduli $G'(\omega)$ and $G''(\omega)$ (Fig. 6c) an oscillation amplitude analysis performed on two different wires of length 15.5 and 15.7 µm provides elastic moduli of 1.2 ± 0.3 Pa and 1.3 ± 0.4. These values are close to the 1.7 ± 0.1 Pa from cone-and plate. Results show here that for viscoelastic solutions with very long relaxation times, magnetic rotation spectroscopy allows a rapid and thorough determination of the rheological parameters.

## IV - Conclusion

We investigate the magnetic wire behaviors to determine the rheological parameters of viscoelastic fluids. In this paper we focus on the rotational magnetic spectroscopy technique that provides the wire motion response over a broad spectral range. Experiments are performed on two wormlike micellar solutions formed by cationic surfactants using an electromagnetic device for wire rotation. The actuating frequency is varied over an extended range, typically between 10$^{-2}$ to 100 rad s$^{-1}$. The surfactant solutions were selected because they are analogues to Maxwell fluids with well-understood structures and dynamics. In particular, the comparison of wormlike micellar solutions with different network relaxation times is emphasized. The main result emerging from the study is that wires dispersed in the micellar fluids exhibit a transition between a synchronous and an asynchronous rotation regime. A wide variety of rotational motions monitored by time-lapse microscopy, including steady rotation, oscillations, continuous and discontinuous back motions is monitored. These behaviors are consistent with mechanistic models developed on the basis of Maxwell fluid environments. The study of the wires angular velocity and oscillation amplitude allows a precise characterization of surfactant solution rheological properties. Viscoelastic fluids presenting relaxation times greater than the half of the inverse critical frequency are shown to display specific sawtooth oscillations in the instability regime, a result that was not reported so far. Our findings pave the way for a rapid and reliable determination of complex fluids rheological responses using the wire-based microrheology technique coupled with a simple data analysis.

## Acknowledgements

The authors are grateful to L. Chevry, A. Cebers, M.-A. Fardin and A. Hallou for fruitful discussions. The master students who participated to the research, C. Leverrier, A. Conte-Daban, N. K. Sampathkumar, C. Lixi, L. Carvhalo and R. Chan are also acknowledged.

## References

[1] R.G. Larson, The Structure and Rheology of Complex Fluids, Oxford University Press, New York, 1998.
[2] H.T. Banks, S. Hu, Z.R. Kenz, A Brief Review of Elasticity and Viscoelasticity for Solids, Adv. Appl. Math. Mech., 3 (2011) 1-51.
[3] T.A. Waigh, Microrheology of complex fluids, Reports on Progress in Physics, 68 (2005) 685-742.
[4] T.M. Squires, T.G. Mason, Fluid Mechanics of Microrheology, Annu. Rev. Fluid Mech., 42 (2010) 413-438.
[5] P. Tierno, Recent advances in anisotropic magnetic colloids: realization, assembly and applications, Phys. Chem. Chem. Phys., 16 (2014) 23515-23528.
[6] L.G. Wilson, A.W. Harrison, A.B. Schofield, J. Arlt, W.C.K. Poon, Passive and Active Microrheology of Hard-sphere Colloids, J. Phys. Chem. B, 113 (2009) 3806-3812.
[7] T.G. Mason, D.A. Weitz, Optical measurements of frequency-dependent linear viscoelastic moduli of complex fluids, Phys. Rev. Lett., 74 (1995) 1250-1253.
[8] F.H.C. Crick, A.F.W. Hughes, The Physical Properties of Cytoplasm - A Study by Means of the Magnetic Particle Method, Exp. Cell Res., 1 (1950) 37-80.

[9] D.B. Allan, D.M. Firester, V.P. Allard, D.H. Reich, K.J. Stebe, R.L. Leheny, Linear and nonlinear microrheology of lysozyme layers forming at the air-water interface, Soft Matter, 10 (2014) 7051-7060.
[10] A. Celedon, C.M. Hale, D. Wirtz, Magnetic Manipulation of Nanorods in the Nucleus of Living Cells, Biophys. J., 101 (2011) 1880-1886.
[11] U. Chippada, B. Yurke, P.C. Georges, N.A. Langrana, A Nonintrusive Method of Measuring the Local Mechanical Properties of Soft Hydrogels Using Magnetic Microneedles, J. Biomech. Eng., 131 (2009) 021014.
[12] P. Dhar, Y.Y. Cao, T.M. Fischer, J.A. Zasadzinski, Active Interfacial Shear Microrheology of Aging Protein Films, Phys. Rev. Lett., 104 (2010).
[13] A. Ghosh, P. Mandal, S. Karmakar, A. Ghosh, Analytical theory and stability analysis of an elongated nanoscale object under external torque, Phys. Chem. Chem. Phys., 15 (2013) 10817-10823.
[14] Y. Gu, Z. Chen, N. Borodinov, I. Luzinov, F. Peng, K.G. Kornev, Kinetics of Evaporation and Gel Formation in Thin Films of Ceramic Precursors, Langmuir, 30 (2014) 14638-14647.
[15] B.H. McNaughton, P. Kinnunen, R.G. Smith, S.N. Pei, R. Torres-Isea, R. Kopelman, R. Clarke, Compact sensor for measuring nonlinear rotational dynamics of driven magnetic microspheres with biomedical applications, J. Magn. Magn. Mater., 321 (2009) 1648-1652.
[16] J.B. Rovner, C.P. Lapointe, D.H. Reich, R.L. Leheny, Anisotropic Stokes Drag and Dynamic Lift on Cylindrical Colloids in a Nematic Liquid Crystal, Phys. Rev. Lett., 105 (2010).
[17] A. Tokarev, B. Kaufman, Y. Gu, T. Andrukh, P.H. Adler, K.G. Kornev, Probing viscosity of nanoliter droplets of butterfly saliva by magnetic rotational spectroscopy, Appl. Phys. Lett., 102 (2013).
[18] L. Zhang, T. Petit, K.E. Peyer, B.J. Nelson, Targeted cargo delivery using a rotating nickel nanowire, Nanomedicine, 8 (2012) 1074-1080.
[19] A. Brasovs, J. Cimurs, K. Erglis, A. Zeltins, J.-F. Berret, A. Cebers, Magnetic microrods as a tool for microrheology, Soft Matter, 11 (2015) 2563-2569.
[20] D. Engstrom, M.C.M. Varney, M. Persson, R.P. Trivedi, K.A. Bertness, M. Goksor, I.I. Smalyukh, Unconventional structure-assisted optical manipulation of high-index nanowires in liquid crystals, Optics Express, 20 (2012) 7741-7748.
[21] B. Frka-Petesic, K. Erglis, J.-F. Berret, A. Cebers, V. Dupuis, J. Fresnais, O. Sandre, R. Perzynski, Dynamics of paramagnetic nanostructured rods under rotating field, J. Magn. Magn. Mater., 323 (2011) 1309-1313.
[22] G. Helgesen, P. Pieranski, A.T. Skjeltorp, Nonlinear phenomena in systems of magnetic holes, Phys. Rev. Lett., 64 (1990) 1425-1428.
[23] G. Helgesen, P. Pieranski, A.T. Skjeltorp, Dynamic behavior of simple magnetic hole systems, Phys. Rev. A, 42 (1990) 7271-7280.
[24] C.W. de Silva, Vibration Damping, Control, and Design, Mechanical and Aerospace Engineering Series, CRC Press Boca Raton, Fla, 2007, pp. 634.
[25] F. Pedaci, Z.X. Huang, M. van Oene, S. Barland, N.H. Dekker, Excitable particles in an optical torque wrench, Nature Phys., 7 (2011) 259-264.
[26] A. Tokarev, A. Aprelev, M.N. Zakharov, G. Korneva, Y. Gogotsi, K.G. Kornev, Multifunctional magnetic rotator for micro and nanorheological studies, Rev. Sci. Inst., 83 (2012).
[27] N. Cappallo, C. Lapointe, D.H. Reich, R.L. Leheny, Nonlinear microrheology of wormlike micelle solutions using ferromagnetic nanowire probes, Phys. Rev. E, 76 (2007) 031505.
[28] R.M. Erb, J. Segmehl, M. Charilaou, J.F. Loeffler, A.R. Studart, Non-linear alignment dynamics in suspensions of platelets under rotating magnetic fields, Soft Matter, 8 (2012) 7604-7609.
[29] R.M. Erb, J. Segmehl, M. Schaffner, A.R. Studart, Temporal response of magnetically labeled platelets under dynamic magnetic fields, Soft Matter, 9 (2013) 498-505.
[30] H. Rehage, H. Hoffmann, Rheological Properties of Viscoelastic Surfactant Systems, J. Phys. Chem., 92 (1988) 4712 - 4719.
[31] H. Rehage, H. Hoffmann, Viscoelastic Surfactant Solutions : Model Systems for Rheological Research, Mol. Phys., 74 (1991) 933 - 973.
[32] J.-F. Berret, G. Porte, Metastable versus unstable transients at the onset of a shear-induced phase transition, Phys. Rev. E, 60 (1999) 4268-4271.
[33] S. Lerouge, J.-F. Berret, Shear-Induced Transitions and Instabilities in Surfactant Wormlike Micelles, in: K. Dusek, J.F. Joanny (Eds.) Polymer Characterization: Rheology, Laser Interferometry, Electrooptics2010, pp. 1-71.
[34] L.M. Walker, P. Moldenaers, J.-F. Berret, Macroscopic response of wormlike micelles to elongational flow, Langmuir, 12 (1996) 6309-6314.
[35] J.-F. Berret, A. Sehgal, M. Morvan, O. Sandre, A. Vacher, M. Airiau, Stable oxide nanoparticle clusters obtained by complexation, J. Colloid Interface Sci., 303 (2006) 315-318.

[36] J. Fresnais, M. Yan, J. Courtois, T. Bostelmann, A. Bee, J.F. Berret, Poly(acrylic acid)-coated iron oxide nanoparticles: Quantitative evaluation of the coating properties and applications for the removal of a pollutant dye, J. Colloid Interface Sci., 395 (2013) 24-30.
[37] J.-F. Berret, Controlling electrostatic co-assembly using ion-containing copolymers: From surfactants to nanoparticles, Adv. Colloids Interface Sci., 167 (2011) 38-48.
[38] M. Yan, J. Fresnais, J.-F. Berret, Growth mechanism of nanostructured superparamagnetic rods obtained by electrostatic co-assembly, Soft Matter, 6 (2010) 1997-2005.
[39] M.M. Tirado, C.L. Martinez, J.G. Delatorre, Comparison of theories for the translational and rotational diffusion-coefficients of rod-like macromolecules - application to short DNA fragments, J. Chem. Phys., 81 (1984) 2047-2052.
[40] L. Chevry, N.K. Sampathkumar, A. Cebers, J.F. Berret, Magnetic wire-based sensors for the microrheology of complex fluids, Phys. Rev. E, 88 (2013) 062306.

# TOC image

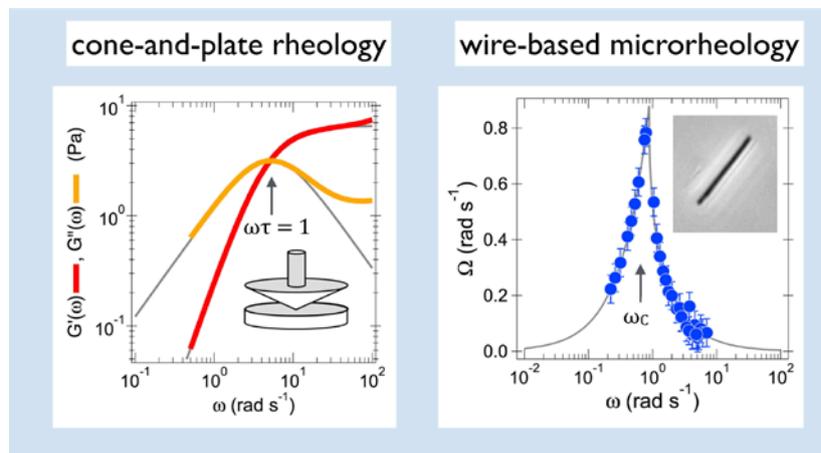